\documentclass[a4paper,12pt]{article}
\usepackage{amsfonts,amsmath,amssymb,appendix}
\usepackage{cite}
\usepackage{url}
\usepackage{hyperref}

\def\Z{\mathbb{Z}}

\def\N{\mathcal{N}}
\def\T{\mathcal{T}}
\def\H{\mathcal{H}}
\def\be{\begin{equation}}
\def\ee{\end{equation}}
\def\nn{\nonumber}
\def\Tr{\textrm{Tr}}
\newcommand{\etabox}[2]{\underset{\ ~#2}{\mbox{\scriptsize $#1$}\ \framebox[15pt]{\phantom{a}}}}

\begin{document}
\begin{titlepage}
\noindent {IITM/PH/TH/2012/1}\\
\bigskip
\begin{center}
\textbf{A non-commuting twist  in the partition function} 
\end{center}
\bigskip
\begin{center}
{Suresh Govindarajan\footnote{\texttt{suresh@physics.iitm.ac.in}} \\
\textit{Department of Physics,\\ Indian Institute of Technology Madras, \\ Chennai 600036, India.} \\
\bigskip
and \\
\bigskip
Karthik Inbasekar\footnote{\texttt{ikarthik@imsc.res.in}}}\\
\textit{Institute of Mathematical Sciences,\\ CIT Campus \\ Taramani, Chennai 600113, India.}
\end{center}
\bigskip
\begin{abstract}

We compute a twisted index for an orbifold theory when the twist 
generating group does not commute with the orbifold group. The twisted 
index requires the theory to be defined on moduli spaces that are 
compatible with the twist. This is carried for CHL models at special 
points in the moduli space where they admit dihedral symmetries. The 
commutator subgroup of the dihedral groups are cyclic groups that are 
used to construct the CHL orbifolds. The residual reflection symmetry is 
chosen to act as a ``twist" on the partition function. The reflection 
symmetries do not commute with the orbifolding group and hence we refer 
to this as a non-commuting twist. We count the degeneracy of half-BPS 
states using the twisted partition function and find that the 
contribution comes mainly from the untwisted sector. We show that the 
generating function for these twisted BPS states are related to the 
Mathieu group $M_{24}$.

\end{abstract}
\end{titlepage}

\newpage
\tableofcontents
\section{Introduction}

The microscopic counting of black hole entropy in four-dimensional 
string theories with $\mathcal{N}=4$ supersymmetry has turned out to 
have a surprisingly rich structure\cite{Dijkgraaf:1996it,Jatkar:2005bh}. 
This has provided connections to modular forms, Lie 
algebras\cite{Govindarajan:2008vi,Cheng:2008kt} as well as sporadic 
groups\cite{Govindarajan:2009qt,Eguchi:2010ej}. Due to the large amount 
of supersymmetry, these theories work as ``laboratories'' for us to test 
ideas that presumably should continue to work in situations with fewer 
super symmetries. This paper seeks to add another variant to the 
microscopic counting -- we `count' twisted half-BPS states in theories 
with $\mathcal{N}=4$ supersymmetry where the twist does not commute with 
the orbifolding group. This is as a prelude to considering situations 
where the twist does not commute with supersymmetry.

We consider four dimensional CHL $\Z_n$-orbifolds with $\N=4$ 
supersymmetry\cite{Chaudhuri:1995fk,Chaudhuri:1995bf} . These models are 
asymmetric orbifolds \cite{Narain:1986qm,Narain:1990mw} constructed by 
starting with a heterotic string compactified on a $T^4 \times S^1 
\times \tilde{S^1}$ and then quotienting the theory by a $\Z_n$ 
transformation which involves a $1/n$ shift along the $\tilde{S^1}$. The 
$\Z_n$ symmetry has a non-trivial action on the internal conformal field 
theory coordinates describing the heterotic compactification on $T^4$.  
A large class of such models where constructed in 
\cite{Chaudhuri:1995dj,Aspinwall:1995fw} and were shown to be dual to a 
type II description compactified on $K3\times S^1 \times \tilde{S^1}$ 
via string-string duality \cite{Sen:1995cj,Harvey:1995rn}.

By construction CHL models possess maximal supersymmetry and fewer 
massless vector multiplets at generic points in the moduli space. The 
requirement of maximal supersymmetry restricts one to consider 
symplectic automorphisms on $K3$. Symplectic automorphisms leave the 
holomorphic $(2,0)$ forms invariant and hence preserve supersymmetry. 
The action of these symmetries have fixed points on the $K3$ surface and 
is accompanied by translations on the circle to avoid quotient 
singularities. So the allowed groups must faithfully represent 
translations in $\mathbb{R}^2$ which implies that the quotienting group 
has to be abelian \cite{Aspinwall:1996mn}. The possible abelian groups 
that act symplectically on $K3$ where classified and the action of the 
group on the $K3$ cohomology was calculated \cite{Nikulin1979}. Once the 
action on the cohomology is determined one uses string-string duality to 
map the action to the Heterotic side. The map is allowed provided the 
supergravity side is free from fixed points, i.e the action on $K3$ must 
be accompanied by shifts on the torus.

The work of Mukai \cite{Mukai1988} opened up the possibility that 
non-abelian groups can act as symplectic automorphisms on the $K3$ 
surface. A couple of years ago Garbagnati \cite{2009arXiv0904.1519G} 
constructed elliptic $K3$ surfaces that admit dihedral group as 
symplectic automorphisms. These automorphisms are constructed by 
combining automorphisms which act both on the base and the fiber such 
that the resulting action is symplectic.  In particular 
\cite{2009arXiv0904.1519G} determined the ranks of the invariant 
sublattice and the orthogonal complement and identified the orthogonal 
complement to the invariant sublattice with the lattices in 
\cite{2008arXiv0806.2753G}. However, for compactifications down to four 
dimensions one cannot quotient by a non-abelian group since these groups 
do not represent translations faithfully. However, one can consider the 
theory to be on special points in the moduli space that admit 
non-abelian symmetries and quotient by the commutator subgroup, which is 
abelian.

In this paper, we consider the CHL $\Z_n$-orbifold models ($3\leq n\leq 
6$ at special points in the moduli space where they admit dihedral 
$D_n=\Z_n\rtimes\Z_2$ symmetry\footnote{In our notation, $D_n$ is the 
dihedral group of order 2n, see section \ref{Non-abelian 
orbifolds}.}.The $\Z_n$ subgroup is the commutator subgroup of the $D_n$ 
group and may be quotiented. The special points in moduli space are 
specified by the elliptic $K3$ surfaces that admit $D_n,3\leq n\leq 6$ 
symmetries constructed in \cite{2009arXiv0904.1519G}. Since the action 
of $\Z_n$ group is known on the $K3$ side, we map it to the heterotic 
string using the string-string duality. We then construct the CHL $\Z_n$ 
orbifold in the heterotic picture and let the additional $\Z_2$ symmetry 
act as a twist in the partition function of the orbifolded theory. These 
twist symmetries are identical to the ones considered in 
\cite{Chaudhuri:1995dj,Aspinwall:1995fw} but without shifts. For $\N=4$ 
supersymmetry to be preserved these twists must commute with all the 
unbroken supersymmetries of the theory. Such twists have been considered 
in the $g\in \Z_n$ twisted partition function \cite{Sen:2009md} for 
unorbifolded theories, which counts the index/degeneracy\footnote{Both 
are identical for the cases considered in this paper.} of elementary 
string states when the theory is restricted to special points in moduli 
space. The $g$-twisted helicity index is defined as 
\be\label{twistedindex} 
B_{2m}^g=\frac{1}{2m!} \Tr[g 
(-1)^{2\ell}(2\ell)^{2m}]\ , 
\ee 
where $g$ generates a symmetry of finite order, $\ell$ is the third 
component of angular momentum of a state in the rest frame, and the 
trace is taken over all states carrying a given set of charges. States 
which break less than or equal to $4m$ $g$-invariant supersymmetries 
give non-vanishing contributions to $B_{2m}^g$ \cite{Sen:2009md}. For 
the case of $1/2$ BPS states that we consider in this paper,
 the relevant index is $B_4^g$.

For our case, the choice of the moduli space that has dihedral symmetry 
is compatible with the $g \in \Z_2$ twist. The other requirement that 
the physical charges have to be $g$ invariant is met by requiring the 
charges $Q$ to take values from lattices invariant under Dihedral 
symmetry \cite{2009arXiv0904.1519G,2008arXiv0806.2753G}. This choice is 
also compatible with the orbifold action, since these lattices possess 
invariance under both $\Z_2$ and $\Z_n$ actions. Thus one meets the 
requirements for the twist and orbifold action to be well defined.

We count the degeneracy of electrically charged $1/2$ BPS elementary 
string states for a fixed charge $Q$ in these theories following the 
method described in \cite{Sen:2005ch}. The $\Z_2$ twisted partition 
function in the $\Z_n$ orbifold theories receives contribution only from 
the orbifold untwisted sector for odd $n$ and additionally from the 
orbifold sector twisted by the element $h^{n/2}$ for even $n$. From the 
point of view of the dihedral group, for even $n$, the element $h^{n/2}$ 
is a nontrivial center of the group and commutes with every element. We 
derive a generating function for these degeneracies and find that it has 
the expected asymptotic limit. The generating function for these twisted 
$1/2$ BPS also forms a Mathieu representation as it did for the abelian 
cases \cite{Govindarajan:2009qt,Govindarajan:2010fu}.

The paper is organised as follows. Following the introductory section, 
in section \ref{Non-abelian orbifolds}, we give a pedagogical 
introduction to non-abelian orbifolds and define the twisted partition 
function to indicate the contributing orbifold twisted sectors. It is 
followed by the construction of CHL $\Z_n$ orbifolds in the heterotic 
picture and the derivation of the half-BPS degeneracies of $g\in \Z_2$ 
twisted BPS states in section \ref{mainsection}. In section 
\ref{Mathieu}, we provide a connection with the sporadic Mathieu group 
$M_{24}$.  We conclude with a summary of our results in section 
\ref{end}.

\section{Non-abelian orbifolds}\label{Non-abelian orbifolds}

In this section we describe the standard CFT approach for constructing 
non-abelian orbifold theories. For a general description of orbifolds in 
string theory see 
\cite{Dixon1985678,Dixon1986285,Dixon198713,philippe1997conformal, 
Ginsparg:1988ui,Dijkgraaf:1989hb,Freed1987}. For some phenomenological 
model building approaches based on non-abelian orbifold string theories 
see \cite{Kakushadze:1996hj,Frampton:2000mq}. Orbifold CFT's are 
generally constructed by considering a theory $\T$ which admits a finite 
discrete symmetry group $G$ consistent with its allowed interactions and 
then forming a quotient theory $\T/G$. When the orbifolding group is 
abelian the entire group acts as the symmetry group whereas in 
non-abelian orbifolds the symmetry group is $G/{[G,G]}$, where $[G,G]$ 
is the commutator subgroup of $G$. As an example, for Dihedral groups 
$D_n$ of order $2n$, the quotienting group is the cyclic group $\Z_n$ of 
order $n$. Before proceeding further, it is useful to define some 
notations. Let us denote the worldsheet coordinate as $X(\tau,\sigma)$, 
with $\tau$ and $\sigma$ being the ``space'' and ``time'' directions of 
the torus. By
\be
\etabox{g}{h}\equiv \Tr_{\H_h}\bigl(g\  q^{H}\bigr)\ ,
\ee
we mean the following closed string boundary conditions are applied 
simultaneously.
\begin{align} 
 X(\tau+2\pi,\sigma)&= g\cdot X(\tau,\sigma) \nn \\
X(\tau,\sigma+2\pi)&= h\cdot X(\tau,\sigma).
\end{align}
$\Tr_{\H_h}$ denotes the trace taken in a Hilbert space sector $\H_h$ 
corresponding to a spatial twist element $h$. $H$ denotes the 
Hamiltonian of the theory. We also denote $|G|$ as the order of the 
group G. The module $\etabox{g}{h}$ is not well defined for $g h\neq h 
g$ as we will explain below.

For the CFT to be well defined, the states of the theory must be 
invariant under the action of the group. Therefore one projects onto 
$G$-invariant states by defining a projection operator
\be
P=\frac{1}{|G|}\sum_{g\in G}g .
\ee
The projection is implemented by including $g$ in the trace and then by 
summing over all time twists.  The inclusion of $g$ in the trace amounts 
to twisting the fields by $g$ along the time direction, i.e $g\cdot 
X(\tau,\sigma)=X(\tau+2\pi,\sigma)$ The contribution to the partition 
function from the spatially untwisted sector of the orbifold CFT is then 
given by
\be
Z_{\H_e}=\frac{1}{|G|}\sum_{g\in G} .
\ee
Modular invariance under $SL(2,\Z)$ transformations requires the 
addition of spatially twisted sectors $\etabox{e}{h}$, i.e sectors where 
fields satisfy $ h\cdot X(\tau,\sigma)=X(\tau,\sigma+2\pi)$. Each of 
these spatially $h$-twisted sectors corresponds to a distinct Hilbert 
space $\H_h$ and one must project onto the group invariant states within 
every Hilbert space. This would mean that the fields would have 
simultaneous boundary conditions due to the action of $g$ and $h$.
\begin{align}
X(\tau,\sigma+2\pi) &= h X(\tau,\sigma) & X(\tau+2\pi,\sigma) &= g X(\tau,\sigma) \nn\\
g X(\tau,\sigma+2\pi)&=g h X(\tau,\sigma) & h X(\tau+2\pi,\sigma) &= g X(\tau,\sigma) \nn\\
g X(\tau,\sigma+2\pi)&=g h g^{-1} g X(\tau,\sigma) &  h X(\tau+2\pi,\sigma) &=h g h^{-1} h X(\tau,\sigma)\nn \\ 
X(\tau+2\pi,\sigma+2\pi) &= g h X(\tau,\sigma) & X(\tau+2\pi,\sigma+2\pi) &= h g X(\tau,\sigma). \label{ibc}
\end{align}
From the above equations, one can see that the action of $g$ takes the 
string in the Hilbert space $\H_h$ to the Hilbert space $\H_{g h 
g^{-1}}$. When $g$ and $h$ do not commute these Hilbert spaces are 
different. The elements $h$ and $h'=g h g^{-1}$ are in the same 
conjugacy class and hence the projection operator would mix Hilbert 
spaces corresponding to elements that belong to a given conjugacy class. 
Thus, one is unable to do a full group invariant projection within the 
Hilbert spaces in the spatially twisted sectors. In the operator 
language, the presence of a time twist $g$ that doesn't commute with the 
spatial twist element $h$ would not allow simultaneous diagonalization 
of their respective matrix representations. Nevertheless one can choose 
a basis for $g$ and it acts on the oscillators and eventually on the 
vacuum. As explained above, the vacuum is not left invariant and the 
vacuum in $\H_h$ taken to the vacuum in $\H_{g h g^{-1}}$.  So the trace 
would be over an off-diagonal matrix with diagonal entries zero and 
hence would vanish. Or equivalently, the path integral vanishes due to 
the inconsistent boundary condition \eqref{ibc}. Since the spatially 
twisted sectors are not invariant under the full group. For a given 
spatially twisted sector $\H_h$ one identifies the little group $N_h$ 
consisting of elements that commute with $h$ and project onto states 
invariant under the little group
\be
Z_{\H_h}= \frac{1}{|N_h|}\sum_{g\in N_h}\etabox{g}{h} .
\ee
The various spatially twisted sectors in a given conjugacy class are 
treated in equal footing and hence the spatially twisted sectors are 
labelled by their conjugacy class $C_i$ instead of the group element 
itself. This follows from ``naive'' modular invariance \footnote{modular 
invariance under $PSL(2,\Z)$ transformations, It is naive because the 
modular transformation $\tau\rightarrow\tau+n$ can introduce anomalous 
phases that could spoil modular invariance}.
\be
Z_{C_i}= \frac{1}{|C_i|}\sum_{h\in C_i} Z_{\H_h} =\frac{1}{|C_i|}\sum_{h\in C_i} \biggl(\frac{1}{|N_h|}\sum_{g\in N_h}\etabox{g}{h}\biggr)
\ee
The group invariant states in the theory are formed by taking a linear 
combination of states from a sector twisted by a group element $g$ and 
all other sectors conjugate to it. The full partition function is then 
given by summing over all the conjugacy classes
\be
Z_{\T/G}=\sum_{C_i} Z_{C_i}
\ee
Since for any group G, the order of the little group $N_h$ is the same 
for every element $h\in C_i$\footnote{this is because every element in a 
conjugacy class has the same order, a group element $h$ is of order $n$ 
if $h^n=1$ } , we have $|G|=|N_h| |C_i|$ for every conjugacy class 
$C_i$. Thus the full CFT partition function for a general non-abelian 
orbifold theory can also be written as
\be\label{non-abelian partition function}
Z_{\T/G}\equiv\frac{1}{|G|}\sum_{\substack{ g,h\in G \\ gh=hg}}\etabox{g}{h} .
\ee

We summarize some properties of Dihedral groups which will be useful 
later. The dihedral group denoted as $D_n$ is of order $2n$. One has the 
presentation,
\be\label{presentation}
D_n \cong\langle h,g | h^n=e, g^2=e,g h g =h^{-1} \rangle.
\ee
where $h$, $g$ generate of $\Z_n$ and $\Z_2$ symmetries respectively.  
The group elements are given by 
$D_n=\{e,h,h^2,\ldots,h^{n-1},g,gh,gh^2,\ldots,gh^{n-1}\}$. The $\Z_2$ 
generator acts as an inversion on the axes of reflection, all the 
elements of the form $gh^j$ are of order 2, i.e $(gh^j)^2=1$. The 
properties of dihedral group depend on whether $n$ is even or odd. For 
odd $n$, $D_n$ has $\lfloor n/2 \rfloor + 2$ conjugacy classes are given 
by (the little groups $N_{c_i}$ for each element $c_i$ in $C_i$ are 
indicated beside)
\begin{align}\label{oddconjugacyclass}
&C_0=\{e\} ;& N_e &= D_n\nn \\ 
&C_1=\{g,gh,gh^2,\ldots,gh^{n-1}\}  ;& N_{c_1} &=\{e,c_1\} \nn\\
&C_k=\{h,h^{n-1}\},\{h^2,h^{n-2}\},\ldots,\{h^{\lfloor n/2 \rfloor},h^{\lfloor n/2 \rfloor+1}\} ; & N_{c_k} &=\Z_n
\end{align}
For even $n$ ,$D_n$ has $n/2+3$ conjugacy classes which are given by
\begin{align}\label{evenconjugacyclass}
 & C_0=\{e\}; & N_e &= D_n \nn\\
 & C_1=\{h^{n/2}\}; & N_{c_1} &= D_n \nn\\
 & C_2=\{g,gh^2,gh^4,\ldots,gh^{n-2}\}; & N_{c_2} &=\{ e,c_2,h^{n/2},c_2 h^{n/2}\}\nn\\
 & C_3=\{gh,gh^3,gh^5,\ldots,gh^{n-1}\};& N_{c_3} &=\{ e,c_3,h^{n/2},c_3 h^{n/2}\} \nn\\
 & C_k=\{h,h^{n-1}\},\{h^2,h^{n-2}\},\ldots,\{h^{n/2-1},h^{n/2+1}\};& N_{c_k} &=\Z_n
\end{align}
The group invariant projection operator for $D_n$ has the property
\begin{align}
 P_{D_n}= & \frac{1}{2n}\biggl(\sum_{j=0}^{n-1}h^j+\sum_{j=0}^{n-1}gh^j\biggr) \nn \\
	= & \frac{1}{2} \sum_{k=0}^{1}g^k\biggl( \frac{1}{n}\sum_{j=0}^{n-1}h^j \biggr)\nn\\
	= & P_{\Z_2}.P_{\Z_n},
\end{align}
which follows from the property of the group elements 
\eqref{presentation}. Even though the element $g$ does not commute with 
elements $h\in \Z_n$, it commutes with the projector of $\Z_n$. Thus if 
we take $g$ to be a twist, it \textit{commutes with the orbifold 
projection}. The $\Z_n$ partition function is given by
\be
Z_{T/\Z_n}=\frac{1}{n}\sum_{j=0}^{n-1}\sum_{k=0}^{n-1}\etabox{h^j}{\ h^k}
\ee
Twisting the partition function by $g\in \Z_2$ amounts to insertion of 
$g$ in the trace,
\be
\Tr_{\H_h}\bigl(g\  q^{H}\bigr)
\ee
By the arguments given in \eqref{ibc} only the following terms 
contribute to the trace,
\be\label{twistedpartitionfunction}
Z^g_{T/\Z_n}=\frac{1}{n}\biggl[\sum_{j=0}^{n-1}\etabox{g h^j}{\ \ e}+\delta_{\frac{n}{2},[\frac{n}{2}]}\sum_{j=0}^{n-1}\etabox{g h^j}{\ \ h^{n/2}}\biggr]
\ee
The second sets of terms are there only for even $n$ as can be seen from 
\eqref{evenconjugacyclass}. We refer to this partition function as the 
``twisted'' partition function. Since the twist generating group $\Z_2$ 
does not commute with the orbifold group $\Z_n$, we refer to it as a 
non-commuting twist. In the following sections, we discuss the orbifold 
action and then evaluate \eqref{twistedpartitionfunction} for the CHL 
$\Z_n$-orbifolds.

\section{Computing the Twisted Partition Function}\label{mainsection}

We adapt the half-BPS counting method of Sen\cite{Sen:2005ch} to compute 
the twisted partition function.  In the notation $D_n=\Z_n\rtimes 
\Z_2=H\rtimes G$, $H$ is the commutator subgroup of $D_n$ which is also 
the orbifolding group. $G$ represents an additional symmetry of the 
theory that appears at special points in the moduli spaces. The CHL 
$\Z_n$-orbifold can be described as an asymmetric orbifold of the 
heterotic string compactified on $T^4 \times T^2$.  The $\Z_n$ symmetry 
acts as a shift on one of the circles in the $T^2$ and as a symmetry 
transformation on the rest of the CFT involving the $T^4$ coordinates 
and the 16 left-moving world-sheet scalars associated with the 
$E_8\times E_8$ gauge group. The action of a group element $h$ of the 
orbifold group $H$ is the combination of a shift $a_h$ and a rotation 
$R_h$ acting on the Narain Lattice $\Gamma^{(22,6)}$. The action of the 
twist $g\in \Z_2$ on the $K3$ side is known 
\cite{Nikulin1979,Chaudhuri:1995dj} and has been used to compute twisted 
indices in \cite{Sen:2009md}. $g$ leaves 14 of the 22 2-cycles of $K3$ 
invariant, in other words it exchanges the two $E_8$'s. Furthermore $g$ 
is not accompanied by shifts. The $g\in \Z_2$ insertion in trace 
requires the physical charges $Q$ to be $g$-invariant and the 
orbifolding requires it to be compatible with the $\Z_n$ orbifold 
projection. Hence, we let $Q$ takes values in the lattices that are 
invariant under $D_n=\Z_n \rtimes \Z_2$ symmetry 
\cite{2008arXiv0806.2753G}. For the rest of the computation we fix the 
value of $Q$. Once this is done the twist $g$ has no further action on 
the lattice.

The set of $R_h\ \forall\ h\in H$ forms a group that describes the 
rotational part of $H$ and is represented as $R_H$. To preserve $\N=4$ 
supersymmetry both $R_H$ and $g$ must act trivially on the right movers. 
In the $K3$ side this is enforced by requiring the respective 
automorphisms to be symplectic. The group $H$ leaves $22-k$ of the $22$ 
left moving directions invariant, where $k$ is the number of directions 
that are not invariant under $H$. Then, $R_H$ can be characterized by 
$k/2$ phases $\phi_j(h)$ with $j=1,2,\ldots,k/2$. The complex 
coordinates $X^j$ represent the planes of rotation and the effect of the 
rotation $R_H$ is to multiply the complex oscillators by phases.

The groups also act on the Narain lattice $\Gamma^{(22,6)}$ and leave a 
sublattice $\Lambda_\perp$ invariant. The orthogonal complement to 
$\Lambda_\perp$ is denoted as $\Lambda_\parallel$. To preserve $\N=4$ 
supersymmetry the right movers take their charge values only from the 
invariant part of the lattices and the non-invariant part of the lattice 
is only due to the $k$ left moving directions that are not invariant 
under the action of the group.. Thus $\textrm{rank}(\Lambda_{\perp 
L})=22-k$ , $\textrm{rank}(\Lambda_\parallel) =k$ and 
$\textrm{rank}(\Lambda_{\perp R})=6.$\footnote{This corresponds to the 
six graviphotons that arise from the toroidal compactification.} The 
total number of $U(1)$ gauge fields in the theory is given by 
$\textrm{rank}(\Lambda_\perp)=22+6-k$. For the $\Z_n$ groups, the values 
of $k$ can be read off from Table \ref{table}.

\begin{table}
\centering
\begin{tabular}{| c | c | c |}
\hline
$G$ & rank$(\Lambda_{\parallel})$ & rank$(\Lambda_{\perp L}) $ \\[0.1cm]
\hline
$\Z_2$ & 8  & 14 \\[0.2cm]
$\Z_3$ & 12 & 10 \\[0.2cm]
$\Z_4$ & 14 &  8 \\[0.2cm]
$\Z_5$ & 16 &  6 \\[0.2cm]
$\Z_6$ & 16 &  6 \\[0.2cm]
$\Z_7$ & 18 &  4 \\[0.2cm]
$\Z_8$ & 18 &  4 \\[0.2cm]
$D_2\simeq\Z_2\times\Z_2$ & 12 & 10 \\[0.2cm]
$\Z_2\times\Z_4$ & 16 &  6 \\[0.2cm]
$\Z_2\times\Z_6$ & 18 &  4 \\[0.2cm]
\hline
\end{tabular}
\caption{For the abelian groups the ranks of the invariant sublattice and  the orthogonal
complement are given in \cite{2008arXiv0801.3992G}.}\label{table}
\end{table}

We recollect some lattice definitions from \cite{Sen:2005ch} for 
convenience. Let $V$ be the $22+6$ dimensional vector space in which the 
Narain lattice $\Gamma^{(22,6)}$ is embedded. The action of a given 
group element $h\in \Z_n$ on $V$ leaves a subspace $V_\perp(h)$ 
invariant. The planes of rotation lie along a subspace denoted as 
$V_\parallel(h)$. It is clear that $V_\parallel(h)$ and $V_\perp(h)$ are 
mutually orthogonal to each other. The action of the entire group thus 
separates the vector space $V$ into an invariant subspace $V_\perp$ and 
its orthogonal complement $V_\parallel$ which are defined 
as\footnote{The sublattice that is invariant under a group $G$ acting on 
a lattice, $\Lambda$, is usually denoted by $\Lambda^G$ and its 
orthogonal complement by $\Lambda_G$.}
\be
V_\perp=\bigcap_{h\in \Z_n}V_\perp(h)\quad ; \quad V_\parallel=\bigcup_{h\in \Z_n} V_\parallel (h)
\ee 
The invariant sublattice $\Lambda_\perp$ and its orthogonal  complement
$\Lambda_\parallel$ are defined as
\be
\Lambda^{\Z_n}:=\Lambda_\perp=\Gamma\bigcap V_\perp \quad ; \quad \Lambda_{\Z_n}:=\Lambda_\parallel=\Gamma\bigcap V_\parallel.
\ee
and
\be
\Lambda_\perp (h)=\Gamma\bigcap V_\perp (h) \quad ; \quad \Lambda_\parallel (h)=\Gamma\bigcap V_\parallel (h)\ ,
\ee
where $\Lambda_\perp (h)$ is the lattice component left invariant by a group
element $h$  and $\Lambda_\parallel (h)$ is the orthogonal complement. 
The ranks of these lattices are the dimensions of their respective vector
spaces.

In the following we describe the heterotic construction of the counting 
\cite{Sen:2005ch} in the untwisted sector as the non-commuting twist 
obtains no contribution from the twisted sectors. The projection is unto 
states invariant under the orbifold group $\Z_n$. For individual 
elements, $h \in \Z_n$ there will be a non-trivial shift vector along 
with the rotation. In order to obtain expressions for $g\in \Z_2$ one 
has to just put the shift vectors $a_g$ to zero. For composite elements 
like $gh$ one has a rotation due to $h$ followed by a reflection on the 
axes of rotation by $g$ and there is also a shift on the lattice due to 
$h$, this follows from the group multiplication law. However one does 
not need such explicit details in the computation as we will show later.

As is known, the momenta and windings in the compact directions of the 
theory takes values in the Narain lattice $\Gamma^{(22,6)}$. The 
(left,right) components of the momentum vector are denoted as 
$\vec{P}=(\vec{P}_L,\vec{P}_R)$. Let $N_L$, $N_R$ be the total level of 
left moving and right moving oscillator excitations respectively. For a 
BPS state, the right movers are kept at the lowest eigenvalue allowed by 
GSO projection, i.e $N_R=0$. The level matching condition in the 
untwisted sector is
\be\label{levelmatch1}
N_L-1+\frac{1}{2}(\vec{P}_L^2-\vec{P}_R^2)=0.
\ee
Let $Q=(\vec{Q}_L,\vec{Q}_R)$ denote the projection of $\vec{P}$ along 
$V_\perp$ and $P_\parallel=(\vec{P}_{\parallel L},0)$ the projection of 
$\vec{P}$ along $V_\parallel$. In an orbifold theory such as this one, 
only the components of $P$ along $V_\perp$ can act as sources for 
electric fields. Since $\N=4$ supersymmetry requires the right-moving 
momenta to take values only from the invariant sublattice, $\vec{P}_R$ 
lies entirely along $V_\perp$, we deduce $\vec{P}_R=\vec{Q}_R$. It is 
then clear that $\vec{P}_L$ has the projection $\vec{Q}_L$ along 
$V_\perp$ and $\vec{P}_{\parallel L}$ along $V_\parallel$. Thus 
$\vec{P}_L$ has a orthogonal decomposition
\be
\vec{P}_L= \vec{Q}_L+\vec{P}_{\parallel L}.
\ee
Writing $N=\frac{1}{2}(\vec{Q}_R^2-\vec{Q}_L^2)$ the level matching 
condition in the untwisted sector \eqref{levelmatch1} reads
\be\label{levelmatch2}
N_L-1+\frac{1}{2}\vec{P}_{\parallel L}^2= N.
\ee
Note that, the information that the charge vector should take values on 
some specific lattice has gone into $N$, and the orbifold projection 
proceeds in the usual way. The counting of the number of 
$\Z_n$-invariant BPS states for a given charge $Q$ is then done by 
implementing the group invariant projection. The contribution to the 
trace with a group element $h\in \Z_n$ inserted comes only from those 
$\vec{P}_{\parallel L}$ which are invariant under the action of $h$, i.e 
from those $\vec{P}_{\parallel L}$ which satisfy the condition
\be
\vec{P}_{\parallel L} \in V_\perp(h).
\ee 
Furthermore, two vectors $P$ and $P'$ in $\Lambda$ which may correspond 
to the same charge vector $Q$ would differ by a constant vector. Hence 
the allowed values of $\vec{P}_{\parallel L}$ for a given charge vector 
$\vec{Q}$ are of the form
\be
\vec{P}_{\parallel L}=\vec{K}(Q)+\vec{p},
\ee 
where $\vec{p}\in \Lambda_\parallel$ and $\vec{K}(Q)\in 
({\Lambda}^{\!*}_\parallel/\Lambda_\parallel)$ is a
constant vector that lies  in the unit cell of $\Lambda_\parallel$.
The total momentum vector can thus be decomposed as 
\be
\vec{P}=\vec{P}_L+\vec{P}_R= (\vec{Q}_L+\vec{P}_{\parallel
L})+\vec{Q}_R=\vec{Q}+(\vec{p}+\vec{K}(Q)).
\ee

When a group element $h$ acts on the vacuum carrying such a momentum 
$\vec{P}$ it will produce a phase \cite{Narain:1986qm}
\be
h\ |P\rangle = e^{2\pi i \vec{a}_h\cdot\vec{Q}} e^{-2\pi i \vec{a}_{h L}\cdot(\vec{p}+\vec{K}(Q))}\ |P\rangle 
\ee
where $\vec{a}_h$ is the shift vector on the lattice associated with the 
group element $h$ and $\vec{a}_{h L}$ is its left moving component. Note 
that there no phases associated with $g$ since $a_g=0$. The negative 
sign is due to the signature of the lattice. Thus we can now express the 
degeneracy of BPS states in the untwisted sector of the orbifold 
carrying a charge $\vec{Q}\in \Gamma_\perp$ as
\begin{align}\label{utdegeneracy}
d(Q) = &\frac{16}{|\Z_n|} \sum_{h\in \Z_n}\sum_{N_L=0}^{\infty}d^{osc}(N_L,h) e^{2\pi i \vec{a}_h\cdot\vec{Q}} \nn\\
&\sum_{\substack{\vec{p}\in\Lambda_\parallel\\ \vec{p}+\vec{K}(Q)\in
V_\perp(h)}} e^{-2\pi i \vec{a}_{h L}\cdot\vec{p}}\
\delta_{N_L-1+\tfrac{1}{2}(\vec{p}+\vec{K}(Q))^2, N}
\end{align}
where $d^{osc}(N_L,h)$ is the number of ways one can construct 
oscillator level $N_L$ from the 24 left-movers weighted by the action of 
$h$. The factor of 16 accounts for the degeneracy of a single BPS 
multiplet. The $\vec{Q}$-dependent phase in the above equation prevents 
us from directly computing the generating function of the degeneracies. 
Sen \cite{Sen:2005ch} evaluates the degeneracy treating $\vec{Q}$ and 
$N$ as independent variables in the right hand side of the above 
equation and calling it $F(Q,\hat{N})$.  Of course, setting 
$\hat{N}=N=\tfrac12 Q^2$ in $F(Q,\hat{N})$, one recovers $d(Q)$.  The 
symbol $\hat{N}$ is used to indicate that $N$ is treated as an 
independent variable.

$F(Q,\hat{N})$ counts the number of states in the CFT which carry a 
given charge $Q$, with right-movers in the ground state. The CFT has 
$\bar{L}_0-L_0$ eigenvalue $\hat{N}-\frac{1}{2} Q^2$ which takes integer 
values from one-loop modular invariance. The integer condition for level 
matching is satisfied only after summing over all the $h$ in the trace. 
A partition function can be defined as follows:
\be\label{generatingfunction}
\tilde{F}(Q,\mu)=\sum_{\hat{N}} F(Q,\hat{N}) e^{-\mu \hat{N}}\ ,
\ee
where $\hat{N}$ runs over values for which $F(Q,\hat{N})$ is non-zero.

$\tilde{F}(Q,\mu)$ acts as a generating function for the degeneracy of 
electrically charged $1/2$ BPS states in the theory. Substituting for 
$F(Q,\hat{N})$ from equation \eqref{utdegeneracy} one obtains
\begin{align}
\tilde{F}(Q,\mu)=\frac{16}{|\Z_n|} &  \sum_{\hat{N}}\bigg[\sum_{h\in
\Z_n}\sum_{N_L=0}^{\infty}d^{osc}(N_L,h) e^{2\pi i \vec{a}_h.\vec{Q}} e^{-2\pi i
\vec{a}_{h L}.\vec{K}(Q)}\nn\\
& \sum_{\substack{\vec{p}\in\Lambda_\parallel\\  \vec{p}+\vec{K}(Q)\in
V_\perp(h)}} e^{-2\pi i \vec{a}_{h L}.\vec{p}}\
\delta_{N_L-1+\frac{1}{2}(\vec{p}+\vec{K}(Q))^2, \hat{N}}\biggr] \
e^{-\mu\hat{N}}\ .
\end{align}
The sum over $\hat{N}$ can be carried out and it gets rid of the 
Kronecker delta function to give
\be\label{partfunc}
\tilde{F}(Q,\mu)=\frac{16}{|\Z_n|}\sum_{h\in \Z_n}e^{2\pi i \vec{a}_h.\vec{Q}} \ 
e^{-2\pi i \vec{a}_{h
L}.\vec{K}(Q)}\ \tilde{F}^{osc}(h,\mu)\ \tilde{F}^{lat}(Q,h,\mu)\ .
\ee
where the oscillator and lattice contribution to the partition function as
\begin{align}
\tilde{F}^{osc}(h,\mu) = & \sum_{N_L=0}^{\infty} d^{osc}(N_L,h) e^{-\mu (N_L-1)}\ ,\nn \\
\tilde{F}^{lat}(Q,h,\mu)= & \sum_{\substack{\vec{p}\in\Lambda_\parallel\\ \vec{p}+\vec{K}(Q)\in V_\perp(h)}} e^{-2\pi i \vec{a}_{h L}.\vec{p}} e^{-\frac{1}{2}\mu(\vec{p}+\vec{K}(Q))^2}\ .
\end{align}
Note that $\tilde{F}^{osc}$ has \textit{no} dependence on $Q$ while 
$\tilde{F}^{lat}$ depends weakly on $\vec{Q}$ only through $\vec{K}(Q)$.

The inverse of the partition function gives the degeneracy
\be\label{integral}
F(Q,\tilde{N})=\frac{1}{2\pi i}\int_{\epsilon-i\pi}^{\epsilon+i\pi} d\mu\ \tilde{F}(Q,\mu)\ e^{\mu \tilde{N}}
\ee
where $\mu=2\pi \tau/i$ and $\epsilon$ is a real positive number. It has 
been argued in \cite{Sen:2005ch} that this integral receives its 
dominant contribution from a small region around the origin. Hence, we 
will take the $\mu\rightarrow 0$ limit later. The oscillator 
contribution is calculated easily by noting that the upon the action of 
a group element $h$ the oscillator acquires a phase $e^{2 \pi\phi_j(h)}$ 
\footnote{Note that the elements $h\in \Z_n$ are of cyclic type, i.e 
$h^n=1$ for some $n\in \Z$, so the phases are all of type $\frac{p}{n}$ 
for some $p\in \Z$}
\be\label{oscillatorcontribution1}
\tilde{F}^{osc}(h,\mu)=q^{-1}\biggl(\prod_{n=1}^\infty 
\frac{1}{1-q^n}\biggr)^{24-k} \prod_{j=1}^{k/2}\biggl(\prod_{n=1}^\infty
\frac{1}{1-e^{2\pi i \phi_j(h)}q^n} \frac{1}{1-e^{-2\pi i \phi_j(h)}q^n}
\biggr),
\ee
where $k$ is the number of non-invariant directions under $\Z_n$. When 
$g$ is inserted into the trace, It will act on the oscillators. The 
phase and number of directions of rotation due to the elements in 
$\tilde{F}^{osc}(h,\mu)$ depends only on the order of the group element. 
In evaluating the oscillator contribution for
\be
\etabox{g}{e}+\etabox{gh}{e}+\etabox{gh^2}{e}+\ldots +\etabox{g h^{n-1}}{\ \ \ e}
\ee
One notices that all the elements $g, gh,\dots, gh^{n-1}$ are of order 
$2$. Hence all of their oscillator contributions are identical to 
$\etabox{g}{e}\nn$. Since $g$ exchanges the $E_8$ co-ordinates, the 
number of directions that are rotated \eqref{table} $k=8$ and non zero 
phases $\phi_j(g)=1/2$. Upon simplification, the oscillator contribution 
becomes
\be\label{oscillatorcontributionoddeven}
\tilde{F}^{osc}(g,\mu)=\frac{1}{\eta(\tau)^8\eta(2\tau)^8}\ ,
\ee
where
\be\label{dedekind}
\eta(\tau)=q^{1/24}\prod_{n=1}^\infty(1-q^n) \quad \text{with} \ q=e^{2\pi i \tau} = e^{-\mu}\ .
\ee
To write down the generating function, we need the lattice contribution 
due a particular group element $h$ which is given by,
\be
\tilde{F}^{lat}(Q,h,\mu)= \sum_{\substack{\vec{p}\in\Lambda_\parallel\\ \vec{p}
+\vec{K}(Q)\in V_\perp(h)}} e^{-2\pi i \vec{a_{h L}}.\vec{p}}
e^{-\frac{1}{2}\mu(\vec{p}+\vec{K}(Q))^2}.
\ee
We have already restricted the charges to take values on the $D_n$ 
invariant lattices, hence $g$ insertion has no further action on the 
lattice. When $h$ is identity the conditions on $\vec{P}_{\parallel L}= 
\vec{P}+\vec{K}(Q) \in V_\perp(h)$ is trivially satisfied since 
$V_\perp(e)=V$. For any other $h$, since we have $\textrm{dim} 
V_\perp(h)< \textrm{dim} (V)$, it follows that
\be
\tilde{F}^{lat}(Q,h,\mu)\leq\tilde{F}^{lat}(\textrm{dim}Q,e,\mu)
\ee
Therefore the dominant contribution is when $h=e$
\be\label{dominantlattice}
\tilde{F}^{lat}(Q,e,\mu)=\sum_{\vec{p}\in \Lambda_\parallel} e^{-\frac{1}{2}\mu(\vec{p}+\vec{K}(Q))^2}\ .
\ee
where the phase has disappeared as the identity element doesn't shift 
the vectors. As mentioned earlier, this lattice theta function depends 
on $Q$, only through the $\vec{K}(Q)\in \Lambda_\parallel^*/\Lambda= 
\Lambda_\perp^*/\Lambda_\perp$ -- thus there are only a finite number of 
lattice sums to consider.

Thus, combining the oscillator and lattice contributions 
\eqref{oscillatorcontributionoddeven} and \eqref{dominantlattice} we get 
the result
\be\label{final}
\boxed{
\tilde{F}(Q,\mu)\sim \frac{16}{|\Z_n|} \frac{\tilde{F}^{lat}(Q,e,\mu)}{\eta(\tau)^8\eta(2\tau)^8}
}\ ,
\ee 
with $\tau =i\mu/2\pi$. The nice thing about the right hand side of the 
above equation is that it depends only on $\vec{K}(Q)$. Thus, up to 
exponentially smaller terms corresponding to $h\neq e$, the right hand 
side is the generating function of $g$-twisted half-BPS states in the 
charge sector $\vec{K}(Q)$. This is the main result of this section.

This $g\in\Z_2$ twisted partition function counts $g$-twisted half-BPS 
states in a $\Z_n$ orbifold theory, so naturally we expect these modular 
forms to have weights smaller than the ones obtained for the untwisted 
orbifold theories. We will check that this is indeed the case by taking 
the asymptotic limit of \eqref{final}. The $\mu\rightarrow 0$ limit of 
Dedekind eta function
\be
\eta(\mu)\simeq e^{-\frac{\pi^2}{6 \mu}}\sqrt{\frac{2\pi}{\mu}}
\ee
and the lattice contribution \eqref{dominantlattice} after doing a 
Poisson resummation is
\be
\tilde{F}^{lat}(e,\mu)\simeq\frac{1}{\textrm{vol}_{\Lambda_\parallel}} \biggl( \frac{\mu}{2\pi} \biggr)^{-\frac{k_{\Z_n}}{2}}\ ,
\ee
up to exponentially suppressed terms. Thus \eqref{final} has 
$\mu\rightarrow 0$ limit
\be
\lim_{\mu\rightarrow 0}\tilde{F}(\mu)\simeq \frac{16}{|\Z_n|}\frac{1}{\textrm{vol}_{\Lambda_\parallel}}e^{2\pi^2/\mu} \biggl(\frac{\mu}{2\pi}\biggr)^{8-\frac{k_{\Z_n}}{2}}
\ee
We compare the weights of the modular forms for the half-BPS states in 
$\Z_n$ orbifolds\cite{Sen:2005ch, Govindarajan:2009qt} and the modular 
forms for $g$ twisted half-BPS states in $\Z_n$ orbifolds
\begin{center}
\begin{tabular}{| c | c | c | c |}
\hline
 Group & $12-\frac{k_{\Z_n}}{2}$ & $8-\frac{k_{\Z_n}}{2}$ & $k_{\Z_n}=$rank$(\Lambda_\parallel)$\\[0.1cm]
\hline
$\Z_3$ & 6 & 2 & 12 \\[0.1cm]
$\Z_4$ & 5 & 1 & 14 \\[0.1cm]
$\Z_5$ & 4 & 0 & 16 \\[0.1cm]
$\Z_6$ & 4 & 0 & 16 \\[0.1cm]
\hline
\end{tabular}
\end{center}
One can see from the above table that the weights for the $g$ twisted 
half-BPS states are indeed smaller.

\subsubsection*{The other contribution for even $n$}

For the even $n$, as noted in the end of section \eqref{Non-abelian 
orbifolds}, we will get additional contribution from the orbifold 
twisted sector due to the element $h^{n/2}$.
\be
\etabox{g}{\ \ h^{n/2}}+\etabox{gh}{\ \ h^{n/2}}+\etabox{gh^2}{\ \ h^{n/2}}+\ldots +\etabox{g h^{n-1}}{\ \ \ h^{n/2}}\ .
\ee
Here again, the oscillator contribution from each module is identical 
since the elements have the same order. The $\Z_n$ groups, for $n$ even 
have $\Z_2$ as a subgroup which would commute with the $g$ twist in the 
partition function to give a $\Z_2 \times \Z_2$.  This case was already 
computed in \cite{Govindarajan:2010fu} (see Appendix A) in the context 
of $\Z_2 \times \Z_2$ and the result is
\be\label{resulteven}
\tilde{F}^{osc}(\mu)=\frac{1}{\ \eta(2\tau)^{12}} \  .
\ee
We need to compute the lattice contribution in this $h^{n/2}$ sector. 
Here $\vec{P}\in \Lambda^{\Z_2}$ the lattice invariant under the $\Z_2$ 
generated by $h^{n/2}$ unlike the untwisted sector where it was in 
$\Lambda$. The charge vectors $Q$ take value in the projection of 
$\vec{P}$ along $V_\perp$. Thus, we have the lattice contribution given 
by
\begin{equation}
\tilde{F}^{lat}(Q,h,\mu)=  \sum_{\substack{\vec{p}\in\Lambda^{\Z_2}_\parallel\\ \vec{p}+\vec{K}(Q)\in V_\perp(h)}} e^{-2\pi i \vec{a}_{h L}.\vec{p}} e^{-\frac{1}{2}\mu(\vec{p}+\vec{K}(Q))^2}\ ,
\end{equation}
where $\Lambda^{\Z_2}_\parallel=\Lambda^{\Z_2}\cap V_\parallel$ and 
$\vec{K}(Q)\in\Lambda^{\Z_2*}_\parallel/\Lambda^{\Z_2}_\parallel$. 
Again, the dominant contribution to the lattice sum occurs when $h=e$. 
The weight of the relevant modular form is now $6-[k/2]$ where $k$ is 
the rank of the lattice $\Lambda^{\Z_2}_\parallel$. We estimate $k$ 
using the relevant cycle shapes for the $\Z_4$ and $\Z_6$ orbifolds to 
be $6$ and $8$ respectively.  when $n=4$, the cycle shape for the 
element $h$ is $1^42^24^4$. The invariant lattice has dimension 
$12=4+2+4$ and thus dim$V_\parallel=24-12=12$. Elements that belong to 
$\Lambda^{\Z_2}_\parallel$ are those that correspond to an 
$h$-eigenvalue equal to $-1$. There are precisely six of them, two 
coming from the two-cycles and four from the four cycles. A similar 
analysis for the cycle shape $1^22^23^26^2$ for $n=6$ shows that each 
three- and six-cycle contribute $2$ elements with $h^3$-eigenvalue equal 
to unity but $h$-eigenvalue not equal to unity and hence $k=8$. A simple 
asymptotic counting as we did earlier then shows that this contribution 
is \textit{larger} than the contribution from the untwisted sector given 
in Eq. \eqref{final}.


\section{Towards Mathieu representations}\label{Mathieu}

We just derived, in the previous section, a formula for the generating 
function of a family of non-commuting twists associated with the 
dihedral symmetry. Eq. \eqref{final} gives the final result of our 
computation. For the case of commuting twists, the counting of 
$\tfrac12$-BPS states saw the appearance of the sporadic Mathieu group, 
$M_{24}$\cite{Govindarajan:2009qt,Govindarajan:2010fu}. It is 
interesting to ask if the final result in Eq. \eqref{final} is 
associated with the Mathieu group. We shall show that this appears to be 
the case. Recall that the answer, up to overall numerical factors, was 
the ratio of a lattice theta function and an eta product. The connection 
with this eta product with the Mathieu group $M_{24}$ was already shown 
in \cite{Govindarajan:2009qt} and thus we need to show that the lattice 
sum has also a relation to $M_{24}$.  Thus, the lattice contribution is 
of the form
\begin{equation}
\Theta^{\parallel}_{\Z_n}(\vec{K}(Q),\tau)\equiv \sum_{\vec{p}\in \Lambda_{\Z_n}} e^{-\frac{1}{2}\mu (\vec{p}+\vec{K}(Q))^2} \ ,
\end{equation}
where $\Lambda_{\Z_n}$ is the orthogonal complement to the $\Z_n$ 
invariant sub lattice of the Lorentzian lattice $\Gamma^{(22,6)}$.

From the work of Nikulin and Mukai\cite{Nikulin1979,Mukai1988}, the 
lattice $\Lambda_{\Z_n}$ has the following properties: 
\begin{itemize} 
\item[(i)] It is a positive definite lattice\footnote{According to the 
sign conventions of Nikulin and Mukai, it should be a negative definite 
lattice -- this translates to a positive definite lattice in our 
convention. Thus the lattice sum is convergent.} with no roots and rank 
$\leq 19$\cite{Nikulin1979}.
\item[(ii)] It can be primitively embedded into a Niemeier lattice(with 
at least one root) and the action of $\Z_n$ can be extended to the full 
Niemeier lattice\cite{Kondo:1998}. This is not quite appropriate for our 
considerations.
\item[(iii)] However, in the appendix to Kondo's paper\cite{Kondo:1998}, 
Mukai refers to unpublished work of his that shows that $\Lambda_{\Z_n}$ 
can also be primitively embedded into the Leech lattice and the action 
of $\Z_n$ can be extended to the full Leech lattice which we denote by 
$L$. One thus has the isomorphism: $\Lambda_{\Z_n} = L_{\Z_n}$, where 
$L_{\Z_n}$ is the orthogonal complement to the $\Z_n$ invariant sub 
lattice of the Leech lattice, $L^{\Z_n}$.
\end{itemize}

This implies that the lattice sum can be mapped to a sum associated with 
a sub lattice of the Leech lattice. It is known that the automorphism 
group of the Leech lattice is the Conway group, $Co_0$. Let $L$ be the 
Leech lattice and $L^G$ the sub lattice that is invariant under the 
action of a finite group $G\subset Co_0$ and $L_G$ its orthogonal 
complement. While $L$ is self-dual, neither $L^G$ nor $L_G$ are 
self-dual. Further, it is not true that $L=L^G\oplus L_G$. Following the 
discussion in the appendix of \cite{Narain:1986qm}, we see that a 
generic lattice vector of $L$ can be written as sum of two lattice 
vectors, one in $L^{G\ast}$ and the other in $L_G^\ast$. Further, for 
every vector in $L^{G^\ast}$, we can associate a vector in $L_G^\ast$ 
modulo $L_G$. Thus, one has an isomorphism $L^{G\ast}/L^G$ and 
$L_G^\ast/L_G$. Putting all this together, we see that a vector in $L$ 
can be identified with two vectors in $L^G$ and $L_G$ along with an 
element of $L^{G\ast}/L^G$. This leads to the following decomposition of 
the theta function associated with the Leech lattice.
\begin{equation}\label{thetadecompose}
\Theta_L(\tau) = \sum_{a\in L^{G\ast }/L^G} \Theta^a_{L^G}(\tau) \Theta^a_{L_G}(\tau)\ ,
\end{equation}
Setting $G=\Z_n\subset M_{23}\subset M_{24}\subset Co_0$, we see that 
$\Theta^a_{L_G}(\tau)$ is the lattice contribution that we computed for 
the non-commuting twist with the label $a$ being identified with 
$\vec{K}(Q)$.

The lattice sums for elements with $\vec{K}(Q)=0$ appear in the 
Monstrous moonshine correspondence of Conway and 
Norton\cite{Conway:1979}. The lattice sums associated with $M_{24}$ have 
been explicitly worked out by Kondo and Tasaka\cite{Kondo:1986a}. The 
$j$-function associated with a conjugacy class, $\rho$, of $M_{24}$ is 
of the form:
\begin{equation}
j_\rho(\tau)=\frac{\Theta_\rho(\tau)}{g_\rho(\tau)}\ ,
\end{equation}
where $g_\rho(\tau)$ are multiplicative eta products associated with the 
conjugacy class, $\rho$, of $M_{24}$\cite{Dummit:1985,Mason:1985}. Kondo 
and Tasaka have given explicit formulae for the theta functions 
$\Theta_\rho(\tau)$ for all $M_{24}$ conjugacy classes. These theta 
functions are lattice sums of sub lattices of the Leech lattice that are 
invariant under the cyclic group generated by the $M_{24}$ group element 
in the conjugacy class $\rho$.  Presumably, their methods can be 
extended to obtain explicit formulae for 
$\Theta^{\parallel}_{\Z_n}(\vec{K}(Q),\tau)$ as well.  We end with the 
comment that $\Theta_\rho(\tau)$ is a modular from of a congruent 
subgroup of the full modular group. Under the action of $S:\tau 
\rightarrow -1/\tau$, it generates the lattice sum associated with 
$L^{G*}$ which shows that $ \Theta^a_{L^G}(\tau)$ can be generated in 
this fashion.

\section{Discussion}\label{end}

In this paper, we have computed generating functions for non-commuting 
$\Z_2$ twists for CHL $\Z_n$ orbifolds ($3\leq n\leq 6$). The generating 
functions turn out be ratios of the theta functions for the $\Z_n$ group 
and eta products associated with the $\Z_2$ group. We have argued that 
the theta functions are also associated with the Mathieu group by using 
an isomorphism that maps the lattices that appear to the sub lattices of 
the Leech lattice. When $n=4$ and $6$, we find additional contributions 
also arise. Our computations did make use of the properties of the 
dihedral group. It would be interesting to extend this method to other 
nonabelian groups as well. On another note, this computation may also be 
extended to $1/4$ BPS states. One can use the symplectic automorphisms 
that act on the elliptic $K3$ directly in the Type IIA theory 
\cite{Sen:2009md}. It will also be useful to consider twists that break 
supersymmetry, which means we would have to consider non-symplectic 
automorphisms on $K3$. Such twists will provide a controlled way to 
count BPS states in $\N=2$ string theories. \\

\noindent \textbf{Acknowledgments:} We thank Shamik Banerjee, Dileep 
Jatkar, Ashoke Sen, Naveen S. Prabhakar and Prasanta Tripathy for 
several useful discussions. K.I. thanks the IMSc string group and NSM 
2011 for giving opportunities to present this work. We also thank Dileep 
Jatkar for his comments on the draft of this manuscript. The work of 
K.I. is supported by the research fellowship of the Institute of 
Mathematical Sciences, Chennai.

\bibliography{nonabeliantwists}
\end{document}